\documentclass[default,iicol,sn-mathphys-num]{sn-jnl}

\usepackage{graphicx}%
\usepackage{multirow}%
\usepackage{amsmath,amssymb,amsfonts}%
\usepackage{amsthm}%
\usepackage{mathrsfs}%
\usepackage[title]{appendix}%
\usepackage{xcolor}%
\usepackage{textcomp}%
\usepackage{manyfoot}%
\usepackage{booktabs}%
\usepackage{algorithm}%
\usepackage{algorithmicx}%
\usepackage{algpseudocode}%
\usepackage{listings}%
\usepackage{caption}%

\theoremstyle{thmstyleone}%

\theoremstyle{thmstyletwo}%

\theoremstyle{thmstylethree}%

\begin{document}

\title {Superconductivity in monolayer-trilayer phase of La$_3$Ni$_2$O$_7$ under high pressure}

\author[1]{\fnm{Chaoxin} \sur{Huang}}
\equalcont{These authors contributed equally to this work.}
\author[1]{\fnm{Jingyuan} \sur{Li}}
\equalcont{These authors contributed equally to this work.}

\author[2]{\fnm{Xing} \sur{Huang}}

\author[1]{\fnm{Hengyuan} \sur{Zhang}}

\author[1]{\fnm{Deyuan} \sur{Hu}}

\author[1]{\fnm{Mengwu} \sur{Huo}}

\author[1]{\fnm{Xiang} \sur{Chen}}

\author[2]{\fnm{Zhen} \sur{Chen}}

\author*[3]{\fnm{Hualei}\sur{Sun}}\email{sunhlei@mail.sysu.edu.cn}

\author*[1]{\fnm{Meng} \sur{Wang}}\email{wangmeng5@mail.sysu.edu.cn}

\affil[1]{\orgdiv{Center for Neutron Science and Technology, Guangdong Provincial Key Laboratory of Magnetoelectric Physics and Devices}, \orgname{School of Physics, Sun Yat-Sen University}, \orgaddress{\city{Guangzhou}, \postcode{510275}, \country{China}}}

\affil[2]{\orgdiv{Beijing National Laboratory for Condensed Matter Physics}, \orgname{Institute of Physics, Chinese Academy of Sciences}, \orgaddress{\city{Beijing}, \postcode{100190},  \country{China}}}

\affil[3]{\orgdiv{School of Science}, \orgname{Sun Yat-Sen University}, \orgaddress{ \city{Shen Zhen}, \postcode{518107}, \country{China}}}

\abstract{The discovery of 80 K superconductivity in pressurized bilayer Ruddlesden-Popper (RP) nickelate La$_3$Ni$_2$O$_7$ has established a new high-temperature superconductor family\cite{Sun2023,Zhang2024h,Hou2023,wang2024b,wang2024p,zhou2025i,Ko2025,Zhou2025a,Osada2025,Hao2025,zhang2025sc,Sakakibara2024t,zhu2024s,li2024sign,Shi2025pr,Nie2025}. The quest to understand the governing principles of RP nickelate superconductivity has become a central focus in condensed matter physics. Here, we report a critical advance by synthesizing and investigating a distinct structural polymorph of the same compound: the monolayer-trilayer (1313) hybrid phase of La$_3$Ni$_2$O$_7$. Under high pressure, synchrotron X-ray diffraction and Raman spectroscopy reveal a structural transition from the orthorhombic $Cmmm$ to the tetragonal $P4/mmm$ space group at 13~GPa. Above 19 GPa, the phase exhibits a clear superconducting transition, confirmed by a zero-resistance state, albeit at a significantly reduced temperature of 3.6 K. The stark contrast with the 80 K transition in the bilayer phase provides a uniquely clean experimental comparison. Our results demonstrate that the superconducting transition temperature is directly governed by the nature of the interlayer coupling, and the bilayer NiO$_6$ block as the essential structural motif for achieving high-$T_\text{c}$ superconductivity in the RP nickelates.}

\maketitle

Since the discovery of superconductivity with a transition temperature $T_\text{c}$ approaching 80 K under pressure in bilayer Ruddlesden-Popper (RP) phase nickelate La$_3$Ni$_2$O$_7$, research on related systems has rapidly attracted widespread attention. The chemical formula of RP phase nickel oxides can be expressed as $A_{n+1}$Ni$_n$O$_{3n+1}$, where $A$ represents a rare-earth cation and $n$ denotes the number of NiO$_6$ octahedral layers in the structure. When $n$ = 2, the bilayer La$_3$Ni$_2$O$_7$ and its doped compounds exhibit the highest known superconducting transition temperature\cite{li2025i, Li2025a}. In this compound, the strong hybridization between Ni-3$d_{z^2}$ and O-2$p_z$ orbitals causes strong superexchange interactions, which play a crucial role in the superconducting pairing\cite{luo2023b,yang2024o,liu2024e,qu2024,Lu2024i1,Sakakibara2024p,Zhang2024str,Jiang2025t,Yang2023i,Lechermann2023}. As $n$ increases, such hybridization weakens due to frustration and $T_\text{c}$ decreases\cite{Qin2024b}. For examples, the trilayer La$_4$Ni$_3$O$_{10}$ and Pr$_4$Ni$_3$O$_{10}$ exhibits a maximum $T_\text{c}$ of approximately 40 K under pressure\cite{zhang2025sc,zhu2024s,li2024sign,Sakakibara2024t,li2024str,huang2024s,zhang2025b}. On the other hand, both $n$ = 2 and 3 compounds similarly undergo structural transitions to the tetragonal phase and suppression of density wave (DW) states prior to the emergence of pressure-induced superconductivity\cite{Khasanov2025p2,Zhang2024h,Zhao2025p,zhang2025b,huang2024s,li2024str}, highlighting the important role of lattice degrees of freedom in the superconductivity of RP phase nickelates.

\begin{figure*}[t]
  \centering
  \includegraphics[width=1.9\columnwidth]{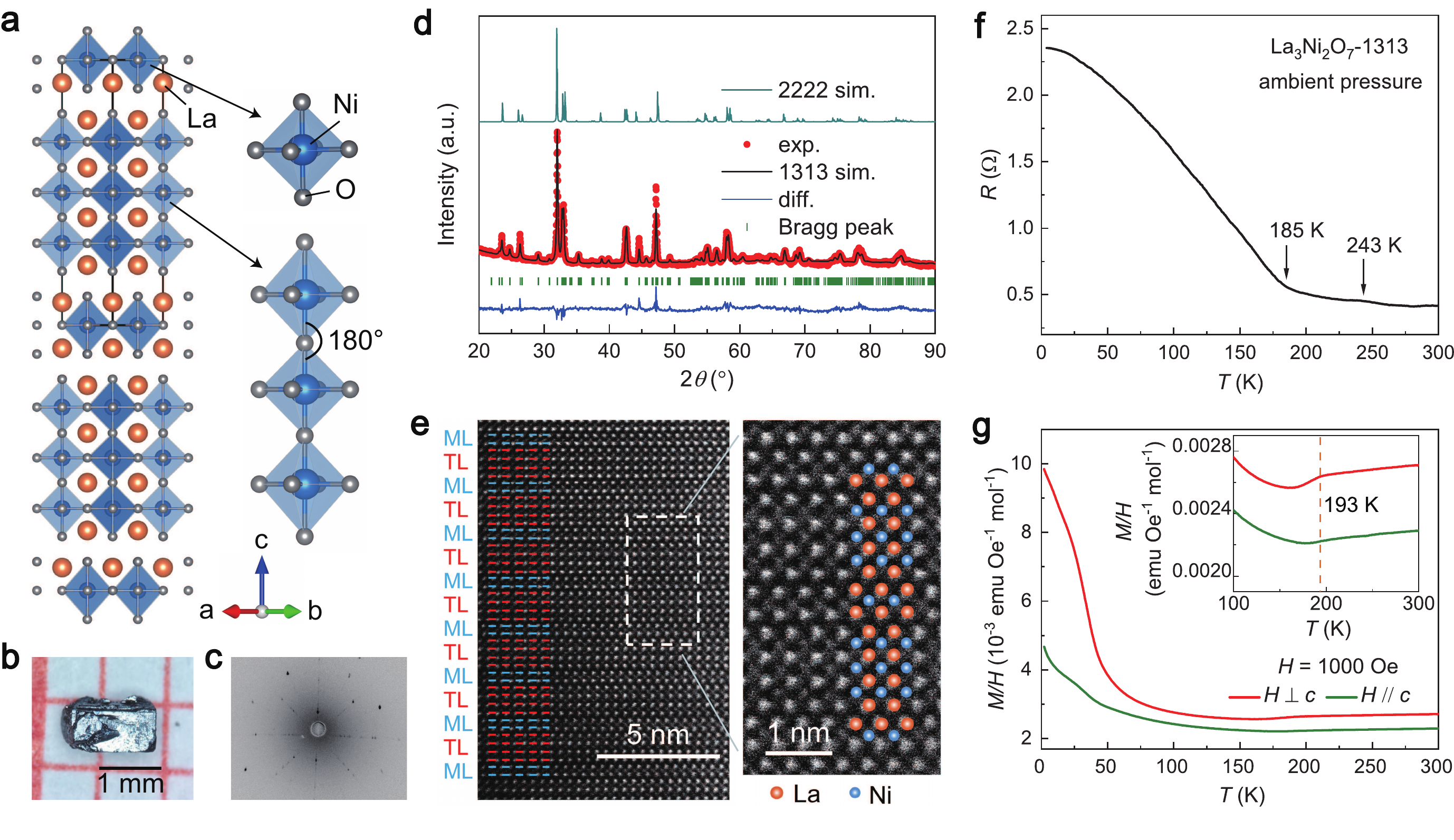}
    \caption{$|$\textbf{Structure, resistance, and magnetization of hybrid La$_3$Ni$_2$O$_7$ at ambient pressure.}
    \textbf{a,} Crystal structure of the monolayer-trilayer (1313) hybrid phase of La$_3$Ni$_2$O$_7$, with an enlarged view illustrating the alternating stacking of monolayer and trilayer NiO$_6$ octahedra. The Ni--O--Ni bond angle along the $c$-axis is 180$^\circ$.
    \textbf{b,c,} Photograph of a single crystal and its corresponding Laue diffraction pattern.
    \textbf{d,} Rietveld refinement of the powder XRD pattern using the 1313 structural model. A simulated pattern for the bilayer (2222) phase of La$_3$Ni$_2$O$_7$ is shown for comparison.
    \textbf{e,} HAADF-STEM image of La$_3$Ni$_2$O$_7$, with a zoomed-in view of the dotted area clearly showing the alternating monolayer (ML) and trilayer (TL) sequences.
    \textbf{f,} Temperature-dependent resistance of a 1313 phase single crystal at ambient pressure, showing two distinct anomalies at 243 K and 185 K.
    \textbf{g,} Temperature-dependent magnetic susceptibility ($M/H$) measured at 1000 Oe for magnetic fields applied in-plane and out-of-plane. The inset shows data from 100 to 300 K, revealing an anomaly at 193 K.}
    \label{fig.1}
\end{figure*}

The hybrid RP phase nickelates have injected new vitality into the research of superconductivity within the nickelate family. To date, two types of hybrid RP phase nickelates have been identified, namely the 1212 phase and the 1313 phase\cite{chen2024p,wang2024lo,puphal2024,li2024de,chen2025v,Shi2025pr,zhang2025e}. The 1212 phase, with the chemical formula La$_5$Ni$_3$O$_{11}$, consists of an alternate stacking sequence of monolayer and bilayer NiO$_6$ octahedral\cite{li2024de}. A pressure-induced superconductivity with the maximum $T_\text{c}$ of 64 K has been reported in this phase\cite{Shi2025pr}. The 1313 phase shares the same chemical formula as the bilayer La$_3$Ni$_2$O$_7$, but its NiO$_6$ octahedral layers are stacked in an alternating monolayer and triayer mode along the $c$ direction\cite{chen2024p}, as shown in Fig.~\ref{fig.1}a. Limited by the purity of the previously measured 1313 phase, conclusive evidence for the superconductivity in the 1313 phase is lacking \cite{puphal2024,Abadi2025,Zhang2024ele2}. Recently, superconductivity has been observed in thin films of the bilayer 2222 phase and the hybrid 1212 phase at ambient pressure. However, the hybrid 1313 thin films grown by the same method do not show superconductivity \cite{Nie2025}.
Given the unique lattice structure of hybrid RP phase nickelates, investigating the superconducting properties of pure 1313 phase of La$_3$Ni$_2$O$_7$ is of significant importance for verifying the superconducting mechanism of both the trilayer and bilayer RP phase nickelates.

In this work, we successfully synthesized high-purity single crystals of the hybrid 1313 phase of La$_3$Ni$_2$O$_7$ using the high-pressure floating-zone method. Our measurements reveal an orthorhombic-to-tetragonal structural transition ($Cmmm$ to $P4/mmm$) at 13~GPa and the emergence of superconductivity above 19~GPa, with $T_\text{c}^{\text{onset}} = 3.6$~K and $T_\text{c}^{\text{zero}} = 2.3$~K at 24.3~GPa. The discovery of superconductivity in the 1313 phase confirms the universality of pressure-induced superconductivity across the RP nickelate family. More importantly, the significantly lower $T_\text{c}$ compared to the bilayer 2222 phase, despite their identical chemical formula, provides compelling evidence that the bilayer NiO$_6$ structural unit itself is a fundamental prerequisite for high-$T_\text{c}$ superconductivity in these materials. Our results establish the number of adjacent, coupled NiO$_6$ planes as a key structural parameter governing superconducting pairing strength, offering a crucial design principle for future materials exploration.

\section*{Structure and properties under ambient pressure}

Figures~\ref{fig.1}b,c show the picture of an as-grown 1313 phase single crystal and its Laue diffraction pattern, which proves a high-quality crystallinity of the single crystal. The Laue diffraction pattern coincides with the (001) plane. X-ray diffraction (XRD) results for the powder ground from single crystals are shown in Fig.~\ref{fig.1}d. The data can be well refined using the 1313 structure model with the orthogonal $Cmmm$ space group, consistent with the previously reported structure\cite{chen2024p,wang2024lo,puphal2024}. Under this symmetry, the angle of the out-of-plane Ni-O-Ni bond within a trilayer NiO$_6$ octahedron maintains 180$^{\circ}$, as shown in the details of Fig.~\ref{fig.1}a. None of the featured peaks of the bilayer phase were detected (Fig.~\ref{fig.1}d). The 1313 alternating stacking structure is also confirmed by high-angle annular dark-field scanning transmission electron microscopy (HAADF-STEM) image shown in Fig.~\ref{fig.1}e. In the enlarged image of the dotted box area, La and Ni cations can be distinguished individually, as marked in Fig.~\ref{fig.1}f. The stacking arrangement is exactly consistent with the 1313 structure shown in Fig.~\ref{fig.1}a.

\begin{figure*}[t]
\centering
  \includegraphics[width=1.9\columnwidth]{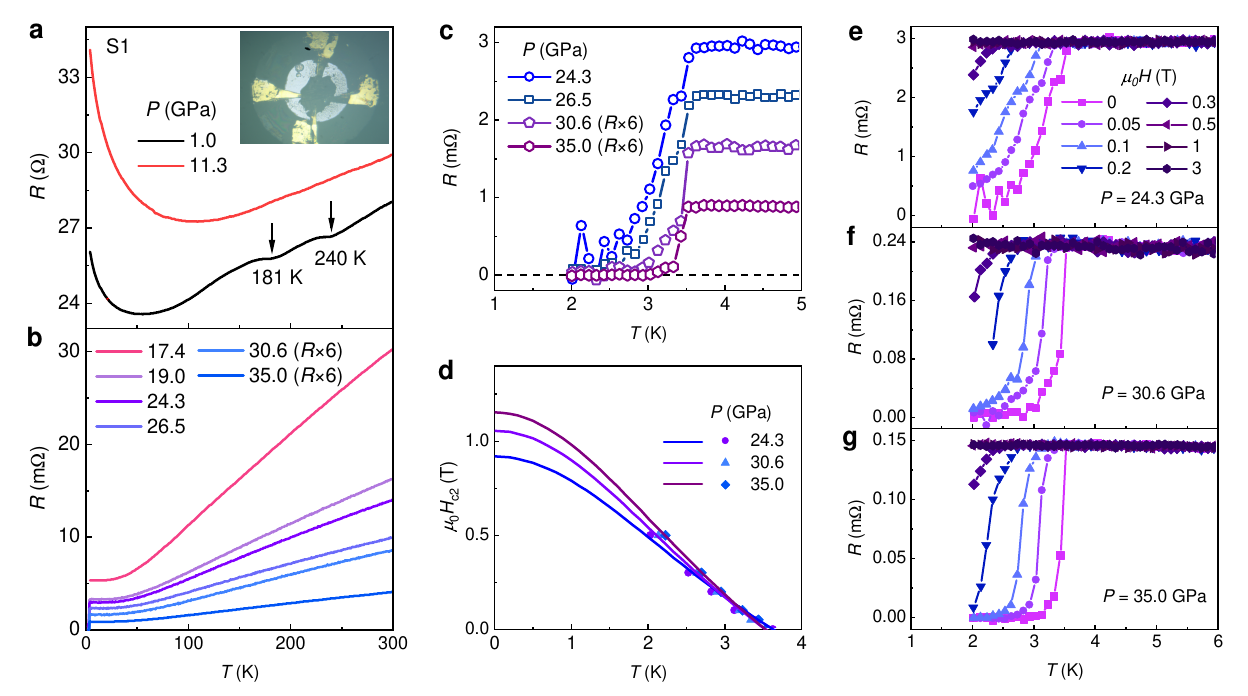}
    \caption{$|$\textbf{Electrical transport properties of hybrid La$_3$Ni$_2$O$_7$ under high pressure.}
    \textbf{a,b,} Temperature-dependent in-plane resistance for sample S1 at various pressures. Two anomalies are visible at 240 K and 181 K at 1.0 GPa. A sharp superconducting transition emerges at $T_\text{c}^\text{onset} = 3.6$ K and 19.0 GPa. The inset shows a photo of the sample configured for high-pressure transport measurements.
    \textbf{c,} Detailed view of the low-temperature resistance of S1 at 24.3, 26.5, 30.6, and 35.0 GPa, demonstrating the development of a zero-resistance superconducting state with increasing pressure.
    \textbf{d,} Temperature dependence of the upper critical field $\mu_0H_{\text{c2}}(T)$ at 24.3, 30.6, and 35.0 GPa, with solid lines representing fits using the Ginzburg-Landau model.
    \textbf{e-g,} Evolution of the superconducting transition with applied magnetic field at selected pressures.}
    \label{fig.2}
\end{figure*}

Temperature dependence of resistance and magnetic susceptibility is measured at 1000 Oe and ambient pressure, as shown in Figs.~\ref{fig.1}f,g. The resistance shows a metallic property but increases slightly at low temperatures. The magnetism susceptibility exhibits significant anisotropy between the in-plane and out-of-plane directions, indicating anisotropic magnetic correlations between Ni cations. The upturn of magnetization below 50 K may be attributed to the minor oxygen vacancies as observed in La$_3$Ni$_2$O$_7$\cite{Liu2023e,Zhang1994}. Two anomalies occur at 243 K and 185 K in resistance. A kink can also be identified at 193 K in magnetic susceptibility, close to the second anomaly in resistance (inset in Fig.~\ref{fig.2}g). The second anomaly is similar to that in other RP phase nickelates such as (La/Pr)$_4$Ni$_3$O$_{10}$\cite{zhu2024s,zhang2025b}, bilayer structural La$_3$Ni$_2$O$_7$\cite{Liu2023e,Zhang2024h}, and the 1212 hybrid La$_5$Ni$_3$O$_{11}$\cite{Shi2025pr}, suggesting a possible spin density wave transition appearing in the 1313 phase. The anomaly at 243 K may originate from the monolayer of NiO$_6$, inherited from the antiferromagnetic order in La$_2$NiO$_4$  \cite{guo1988elec}.

\section*{Pressure-induced superconductivity}

Figure \ref{fig.2} presents high-pressure electrical transport measurements on single crystals (S1 and S2) of the 1313 phase of La$_3$Ni$_2$O$_7$. At 1.0~GPa, two subtle kinks are observed around 240~K and 181~K, consistent with the anomalies measured at ambient pressure (Fig.~\ref{fig.1}f); these features are suppressed by a pressure of 11.3~GPa. A sharp drop in resistance emerges below $T_\text{c}^{\text{onset}} = 3.6$~K at 19.0~GPa (Fig.~\ref{fig.2}b). This drop evolves into a clear zero-resistance state below $T_\text{c}^{\text{zero}} = 2.3$~K at 24.3~GPa, confirming the emergence of superconductivity (Fig.~\ref{fig.2}c). While $T_\text{c}^{\text{onset}}$ remains relatively constant, $T_\text{c}^{\text{zero}}$ increases to 3.2~K at 35.0~GPa. To further verify the superconducting state, we measured the field-dependent resistance at 24.3, 30.6, and 35.0~GPa (Figs.~\ref{fig.2}e--g). The superconductivity is completely suppressed by a magnetic field of 0.5~T at 2~K. The upper critical field $\mu_0 H_{\text{c2}}(0)$, estimated from Ginzburg-Landau fitting, is approximately 1~T across these pressures (Fig.~\ref{fig.2}d).

For type-$\mathrm{II}$ superconductors, the upper critical field is given by $\mu_0 H_{\mathrm{c2}}(T) = \Phi_0 / 2\pi\xi^2(T)$, where $\Phi_0$ is the magnetic flux quantum and $\xi(T)$ is the superconducting coherence length. At 35.0~GPa, we obtain a coherence length $\xi(0) \approx 169$~\AA. Given that $\xi(T) \propto \upsilon_F / \Delta$, where $\upsilon_F$ is the Fermi velocity and $\Delta$ is the superconducting gap, the observed increase in $\mu_0 H_{\mathrm{c2}}$ from 24.3 to 35.0~GPa suggests a moderate enhancement of the pairing gap. This occurs in pressurized superconducting La$_3$Ni$_2$O$_{7-\delta}$ with different oxygen stoichiometries, where $T_\mathrm{c}^{\mathrm{onset}}$ remains nearly constant but the $\mu_0 H_{\mathrm{c2}}$ is quite different\cite{Ueki2025}.

\begin{figure*}[t]
\centering
  \includegraphics[width=1.9\columnwidth]{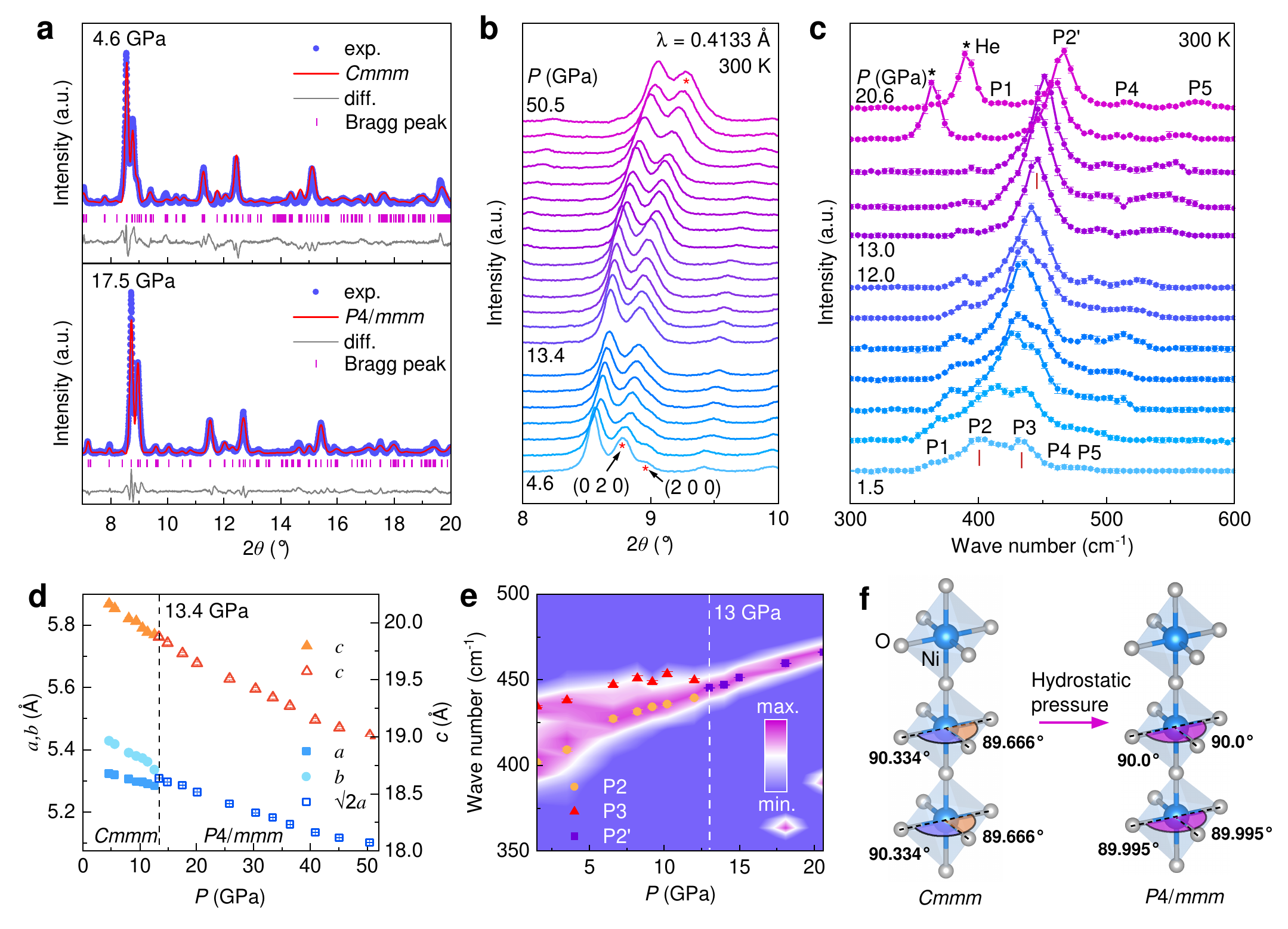}
    \caption{\label{fig.3}$|$\textbf{High-pressure structural evolution of hybrid La$_3$Ni$_2$O$_7$.}
    \textbf{a,} Rietveld refinements of synchrotron XRD patterns for the 1313 phase of La$_3$Ni$_2$O$_7$ at 4.6 GPa and 17.5 GPa, measured at room temperature. The refinements correspond to the orthorhombic $Cmmm$ and tetragonal $P4/mmm$ space groups, respectively.
    \textbf{b,} Pressure evolution of selected reflection peaks between 8$^\circ$ and 10$^\circ$, showing the merging of the (0~2~0) and (2~0~0) peaks, which signals the transition to the tetragonal phase. Neon was used as the pressure-transmitting medium.
    \textbf{c,} Raman spectra of the 1313 phase at room temperature under various pressures. Helium was used as the pressure-transmitting medium.
    \textbf{d,} Pressure dependence of the lattice constants $a$, $b$, and $c$.
    \textbf{e,} Color map of the Raman intensity as a function of pressure and wavenumber. The positions of peaks P2, P3, and P2$^\prime$, determined by Lorentzian fits, are overlaid. The intensity of peak P2 is normalized to unity at each pressure for comparison. The dashed line at 13 GPa marks the merging of two peaks, indicating a structural transition.
    \textbf{f,} Schematic of the trilayer structure unit in the 1313 phase before and after the structural transition.}
\end{figure*}

Similar high-pressure electronic transport measurements were performed on a second sample, S2 (Extended Data Fig. 1). The data from S2 exhibit improved metallic behavior at low pressures and indicate the emergence of a superconducting state below approximately 3.0 K at pressures above 24.5 GPa. However, the resistance does not drop to zero, which may suggest sample inhomogeneity, a phenomenon also observed in other bilayer and trilayer RP nickelate phases\cite{Chen2025l,Zhang2024eff,huang2024s}. Notably, no signature of superconductivity near 80~K was detected in either sample up to the maximum pressure of 35.0~GPa in our experiments.

Among the superconducting RP nickelates, the 1313 phase exhibits the most fragile superconductivity. The maximum ratio $\mu_0H_{\text{c2}}(0)/T_\text{c} \sim 0.33$ is significantly lower than that of bilayer ($\sim$1.5--2.5) and trilayer ($\sim$1.5--1.8) single crystals~\cite{Sun2023,li2025i,Li2025a,zhang2025b,zhang2025sc,zhu2024s}. Interestingly, this value is comparable to the ratio of $\sim$0.44 observed in the 1212 hybrid phase~\cite{Shi2025pr}. The substantially reduced $\mu_0H_{\text{c2}}(0)$ relative to $T_\text{c}$ in both hybrid phases suggests that superconducting electron pairing is confined to the coupled NiO$_6$ layers. The interstitial monolayer NiO$_6$ octahedra, analogous to pressurized La$_2$NiO$_4$, do not contribute to superconductivity~\cite{Ji2024}.

\section*{Structural transition under high pressure}

To investigate the structural evolution under pressure, we performed synchrotron XRD and Raman spectroscopy measurements. Figure~\ref{fig.3} summarizes the high-pressure powder XRD patterns and Raman spectra acquired at room temperature. The XRD pattern at 4.6~GPa is consistent with the ambient-pressure $Cmmm$ structure (Fig.~\ref{fig.3}a), in agreement with the as-grown single-crystal structure~\cite{chen2024p,wang2024lo,puphal2024}. At 13.4~GPa, the adjacent (0~2~0) and (2~0~0) diffraction peaks merge (Fig.~\ref{fig.3}b), indicating a transition to a higher-symmetry tetragonal phase. This transition is further supported by the pressure dependence of the peak positions (Extended Data Fig.~2). Rietveld refinement confirms that the high-pressure phase is well-described by the tetragonal $P4/mmm$ space group (Fig.~\ref{fig.3}a), consistent with a previous high-pressure study~\cite{puphal2024}. The pressure evolution of the lattice constants and the in-plane Ni--O--Ni angle are summarized in Figs.~\ref{fig.3}d,f, whose raw data are shown in Extended Data Figs. 2. The precise determination of the structural transition pressure from XRD is challenging due to the weak X-ray scattering cross-section of oxygen atoms.

Figure~\ref{fig.3}c displays the high-pressure Raman spectra. At 1.5~GPa, five distinct peaks are observed in the 300--600~cm$^{-1}$ range. With increasing pressure, the two prominent peaks P2 and P3 gradually converge, eventually merging into a single peak, P2$^\prime$. The intensity and position of these peaks reveal a structural transition at approximately 13.0~GPa (Fig.~\ref{fig.3}e). Combined, the XRD and Raman measurements demonstrate that the orthorhombic-to-tetragonal structural transition occurs at 13.0 GPa, preceding the emergence of superconductivity at 19 GPa. The Ni--O--Ni bond angle along the $c$-axis has been proposed as a critical parameter for superconductivity in RP nickelates. In the orthorhombic $Cmmm$ structure of the 1313 phase La$_3$Ni$_2$O$_7$ at ambient pressure, this angle is 180$^\circ$, satisfying the proposed criterion. The separation between the structural and superconducting transition pressures in the hybrid 1313 and 1212 phases, in contrast to their coincidence in the bilayer and trilayer phases, highlights the crucial role of the 180$^\circ$ Ni--O--Ni bond angle along the $c$-axis in the emergence of superconductivity in RP nickelates~\cite{Shi2025pr,li2025i,zhu2024s}.

\section*{Phase diagram of hybrid La$_3$Ni$_2$O$_7$}

\begin{figure}[t]
\centering
  \includegraphics[width=1\columnwidth]{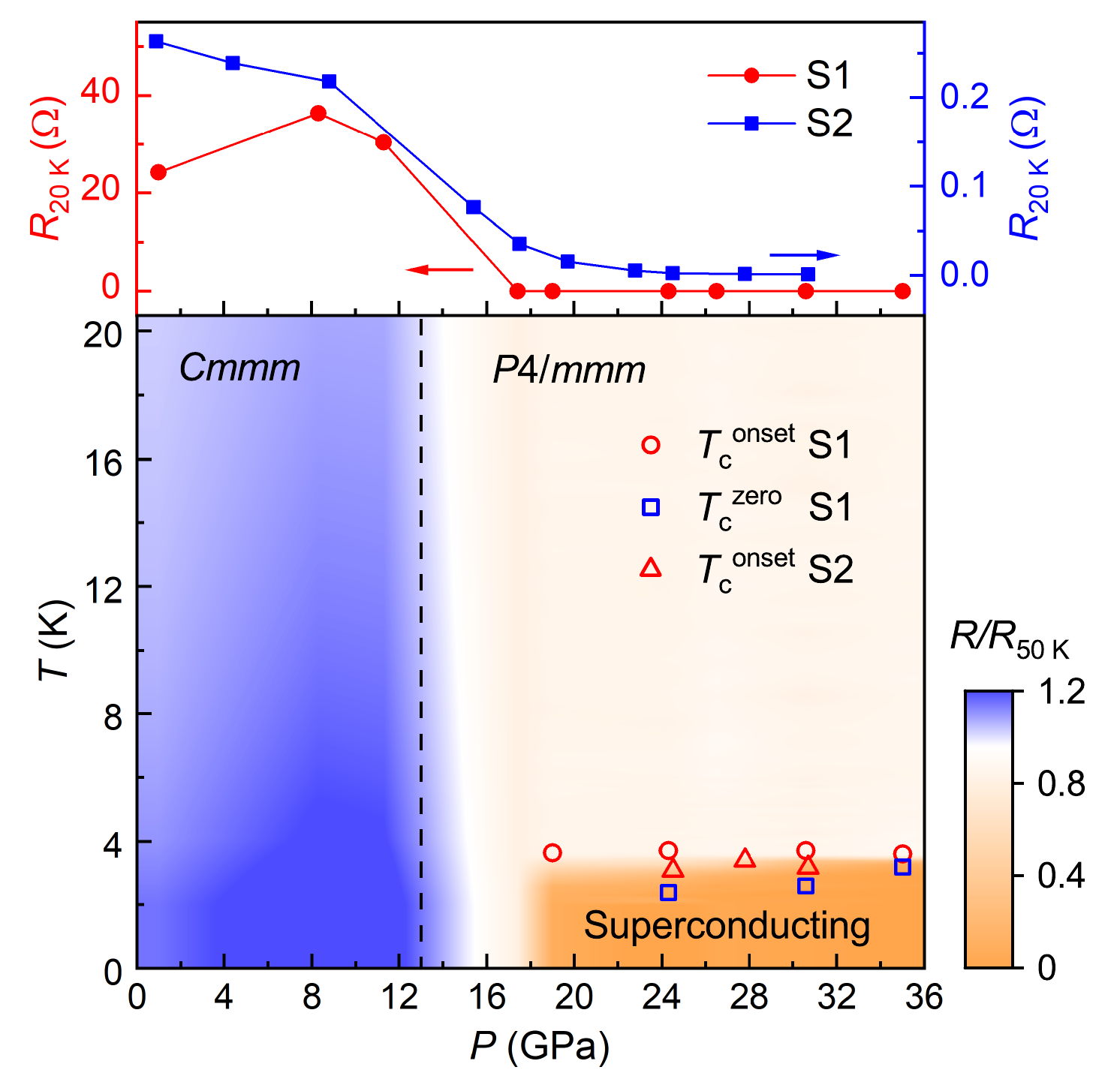}
    \caption{\label{fig.4}$|$\textbf{Phase diagram of hybrid La$_3$Ni$_2$O$_7$.}
  \textbf{Upper panel:} Pressure dependence of the resistance at 20~K for samples S1 and S2.
\textbf{Lower panel:} Temperature-pressure phase diagram for the 1313 phase of La$_3$Ni$_2$O$_7$. The superconducting onset temperature $T_\text{c}^\text{onset}$ and zero-resistance temperature $T_\text{c}^\text{zero}$ for S1, along with $T_\text{c}^\text{onset}$ for S2, are indicated. The pressure-induced structural transition from the orthorhombic $Cmmm$ phase to the tetragonal $P4/mmm$ phase at 13~GPa is marked by a vertical dashed line. The background color represents the normalized resistance $R/R_\text{50 K}$ for S1, highlighting the superconducting region and the pronounced resistance change at the structural transition.}
\end{figure}

We summarize the electronic transport and structural properties of the 1313 phase of La$_3$Ni$_2$O$_7$ from 0 to 36 GPa in the phase diagram of Fig.~\ref{fig.4}, where the color scale represents the normalized resistance $R/R_{50\text{K}}$. The orthorhombic-to-tetragonal structural transition is accompanied by a substantial change in resistance, demonstrating a strong coupling between the electronic and lattice degrees of freedom. Superconductivity in the 1313 phase emerges within the tetragonal structure, mirroring its trilayer counterpart. A systematic comparison across the RP nickelate family reveals a dramatic hierarchy in superconducting transition temperatures: from the bilayer phase ($T_\text{c} \sim 80$~K) to the trilayer phase ($T_\text{c} \sim 30$~K), and further to the hybrid 1212 ($T_\text{c} \sim 64$~K) and 1313 ($T_\text{c} \sim 3$~K) phases. Structurally, the trilayer can be viewed as a frustrated bilayer, where the non-bonding Ni-3$d_{z^2}$ state introduces frustration, weakening the correlations established by the bonding and antibonding states. The 1212 phase is a separated bilayer, and the 1313 phase is both frustrated and separated. Both the bilayer and trilayer blocks of the hybrid phases are compressed in-plane and stretched out-of-plane more than their counterparts (Extended Data Fig. 3). The pronounced suppression of $T_\text{c}$ from the bilayer to the trilayer structure, and to their hybrid phases, underscores the critical role of interlayer electronic and magnetic exchange couplings. Our results, therefore, establish the pristine, coupled bilayer as the essential structural motif for achieving high-$T_\text{c}$ superconductivity in RP nickelates, providing a fundamental principle for guiding the search for new superconducting materials.

\backmatter

\section*{Methods}

\subsection*{Sample synthesis}

Single crystals of the 1313 phase of La$_3$Ni$_2$O$_7$ were grown using the high-pressure floating-zone method. Feed and seed rods were prepared by mixing pre-dried La$_2$O$_3$ (99.99\%) and NiO (99.99\%) powders in a molar ratio of $3:4$. The mixture was ground, pressed into pellets, and sintered at 1100~$^{\circ}$C for 48 h. This process of grinding and sintering was repeated 2--3 times. The resulting powder was then ground again, pressed into rods, and sintered at 1200~$^{\circ}$C for 72 h to form feed and seed rods with a length of 10 cm and a diameter of 5 mm.

Crystal growth was performed in a vertical optical floating-zone furnace (HKZ, SciDre) under an oxygen pressure of 15~bar. The feed and seed rods were counter-rotated at a relative speed of 35 rpm, with a growth rate of 3 mm/h. Throughout the growth process, the molten zone width was minimized. The resulting crystal was approximately 3~cm in length. Pure-phase La$_3$Ni$_2$O$_7$-1313 single crystals were obtained.

\subsection*{Sample characterization at ambient pressure}

The crystal structure of the 1313 phase of La$_3$Ni$_2$O$_7$ at ambient pressure was characterized by XRD (Rigaku), Laue diffraction (Photonic Science), and HAADF-STEM. HAADF-STEM experiments were conducted using a double aberration-corrected JEOL ARM200F microscope operated at 200~kV. DC magnetic susceptibility and electrical resistance measurements were performed using a Superconducting Quantum Interference Device (SQUID) and a Physical Property Measurement System (PPMS), respectively.

\subsection*{High-pressure electronic transport, X-ray diffraction, and Raman measurements}

High-pressure experiments were conducted using DACs. For electronic transport measurements, a DAC with 300-$\mu$m culet anvils was used. Single crystals of dimensions $80 \times 80 \times 10~\mu$m were loaded into a sample chamber of 160~$\mu$m diameter and approximately 30~$\mu$m thickness. Daphne oil 7373 was used as the pressure-transmitting medium (PTM) to ensure a hydrostatic pressure environment. Resistance measurements were performed using a PPMS.

High-pressure synchrotron XRD experiments were conducted at the ID31 High-pressure Beamline of the Beijing High Energy Photon Source. Powder samples, obtained by grinding single crystals, were loaded into a DAC with 250-$\mu$m culet anvils using neon gas as the PTM.

High-pressure Raman spectroscopy measurements employed a DAC with 250-$\mu$m culet anvils and helium gas as the PTM. A single crystal, measuring $50 \times 50 \times 10~\mu$m, was loaded for measurement. Raman spectra were collected in a confocal geometry (MoniVista CRS3) using a Princeton Instruments HRS-500 spectrometer. A 532-nm laser with a power of less than 3 mW was used for excitation. Pressure calibration below 10.0 GPa was performed using the fluorescence spectrum of a 10-$\mu$m ruby sphere; for pressures above 10.0 GPa, the high-frequency edge of the diamond Raman spectrum was utilized.

\bmhead{Acknowledgements}
This work was supported by the National Natural Science Foundation of China (Grants No. 12425404, 12494591, 12474137, 92565303, 52273227, and U22A6005), the National Key Research and Development Program of China (Grants No. 2023YFA1406000, 2023YFA1406500), the Guangdong Basic and Applied Basic Research Funds (Grant No. 2024B1515020040, 2025B1515020008, 2024A1515030030), the Shenzhen Science and Technology Program (Grants No. RCYX20231211090245050), the Guangzhou Basic and Applied Basic Research Funds (Grant No. 2024A04J6417), the CAS Superconducting Research Project (Grant No. SCZX-0101), the Guangdong Provincial Key Laboratory of Magnetoelectric Physics and Devices (Grant No. 2022B1212010008), and Research Center for Magnetoelectric Physics of Guangdong Province (Grant No. 2024B0303390001). High-pressure synchrotron X-ray measurements were performed at the ID31 High-pressure Beamline of the High Energy Photon Source. The ID31 High-pressure Laboratory supported gas loading. We thank Jinlong Zhu from the Department of Physics at the Southern University of Science and Technology for supporting the gas loading for the Raman measurement.

\bmhead{Author contributions}
M.W. and H.L.S. supervised the research. C.X.H., D.Y.H., and M.W.H. grew and characterized the single crystals. J.Y.L., H.Y.Z., and X.C. performed the high-pressure measurements and analyzed the data. X.H. and Z.C. conducted the STEM measurements.

\bmhead{Competing interests}
The authors declare no competing interests.

\bibliography{reference}

\begin{figure*}[t]
\centering
  \includegraphics[width=1.7\columnwidth]{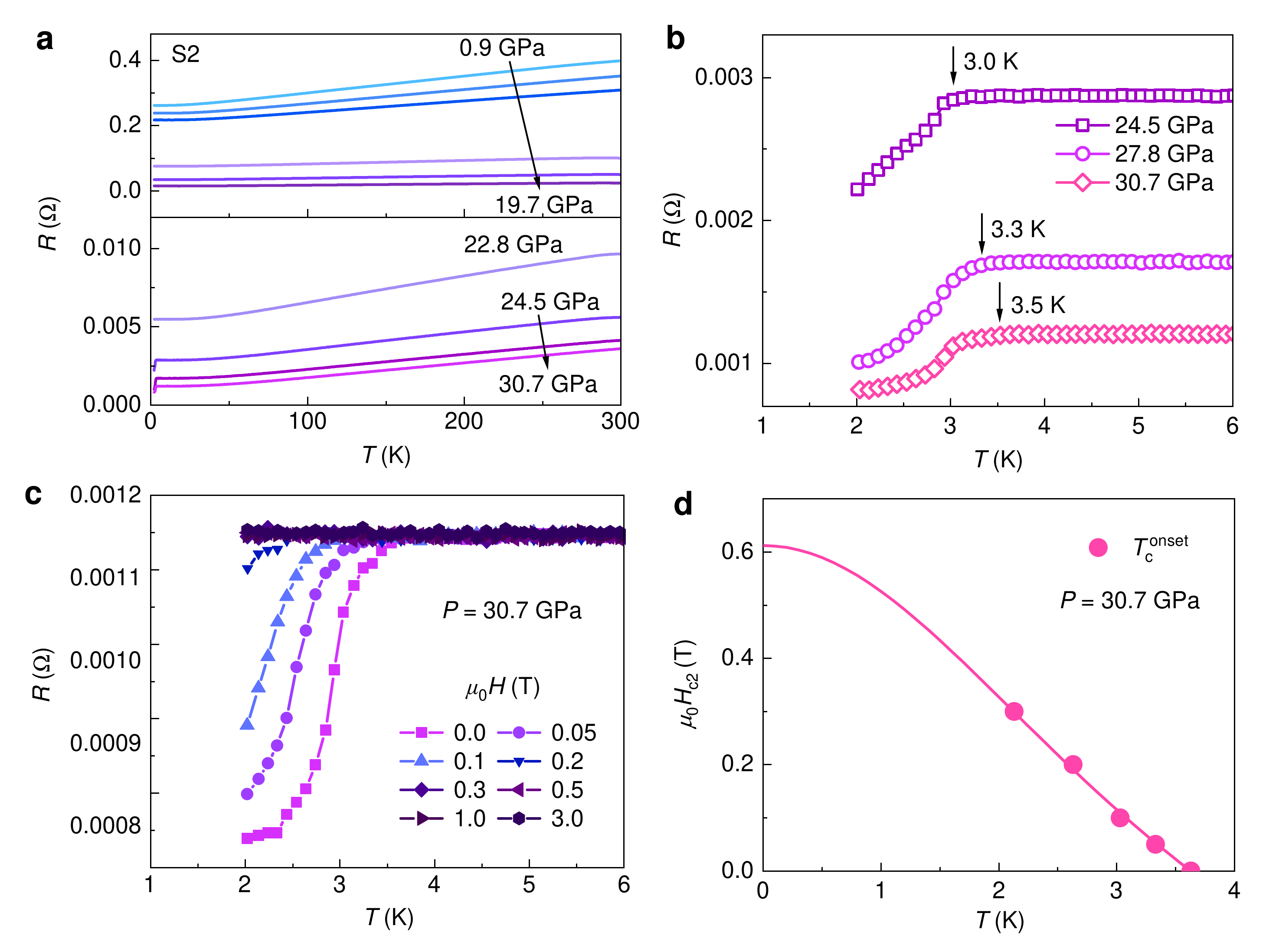}
    \caption*{\textbf{Extended Data Fig.1$|$High-pressure electrical transport properties of S2.}
    \textbf{a,} Temperature dependence of resistance at pressures from 0.9 to 30.7 GPa.
    \textbf{b,} Enlarged display of the resistance at selected pressures below 6 K.
    \textbf{c,} Temperature dependence of the resistance under various magnetic fields below 6 K.
    \textbf{d,} Upper critical field $\mu_0 H_{\text{c2}}(T)$ at 30.7~GPa with a fit based on the Ginzburg-Landau model.}
\vspace*{6in}
\end{figure*}

\begin{figure*}[t]
\centering
  \includegraphics[width=1.7\columnwidth]{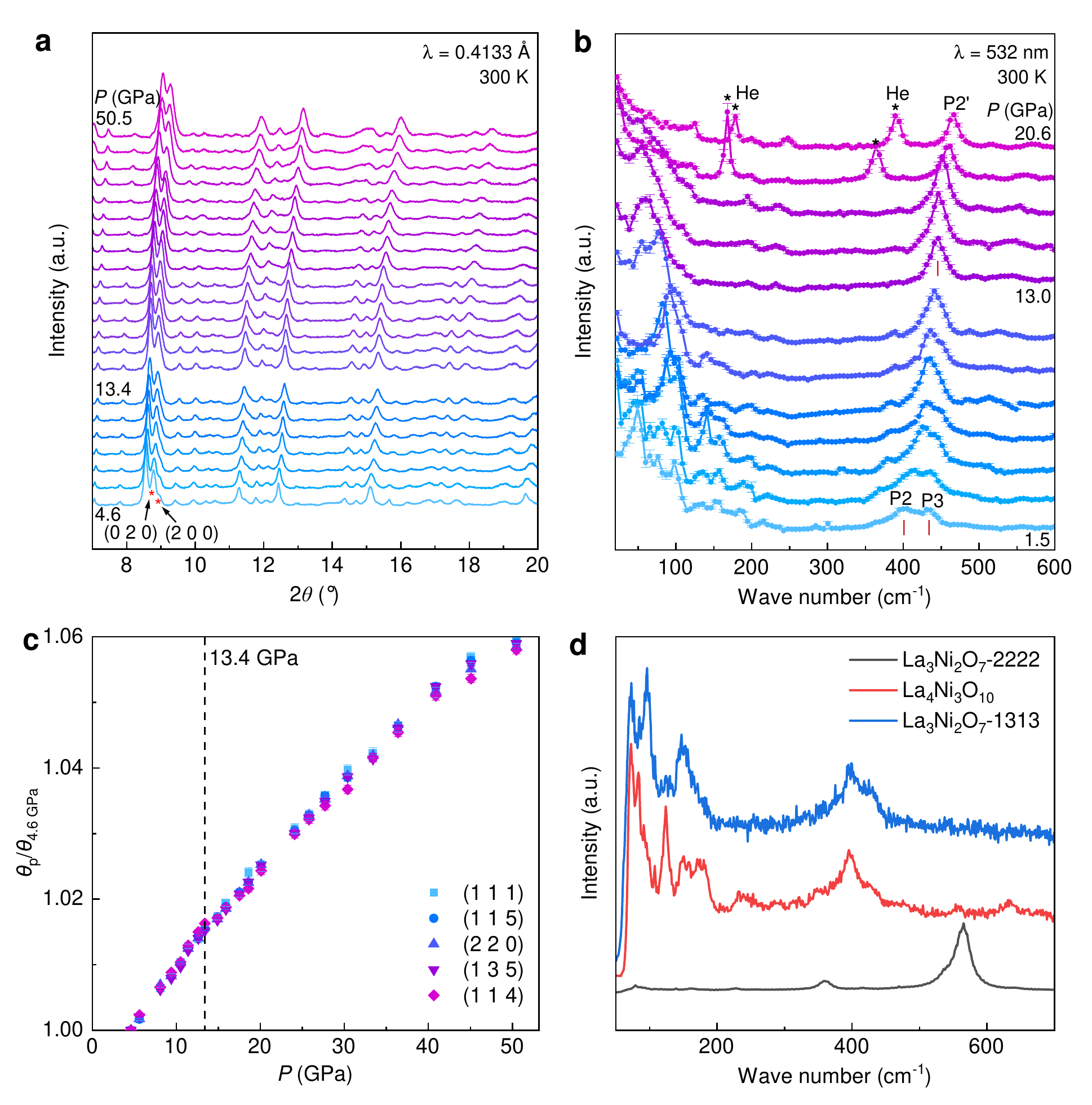}
    \caption*{\textbf{Extended Data Fig.2$|$High-pressure XRD pattern and Raman spectrum of hybrid La$_3$Ni$_2$O$_7$.} \textbf{a,b,} Raw data of the high-pressure XRD pattern and Raman spectrum. \textbf{c,} Normalized peak positions evolution under pressure. \textbf{d,} Comparison of the Raman spectra of the bilayer 2222 phase of La$_3$Ni$_2$O$_7$, trilayer La$_4$Ni$_3$O$_{10}$, and the hybrid 1313 phase of La$_3$Ni$_2$O$_7$. }
\vspace*{6in}
\end{figure*}

\begin{figure*}[t]
\centering
  \includegraphics[width=1.7\columnwidth]{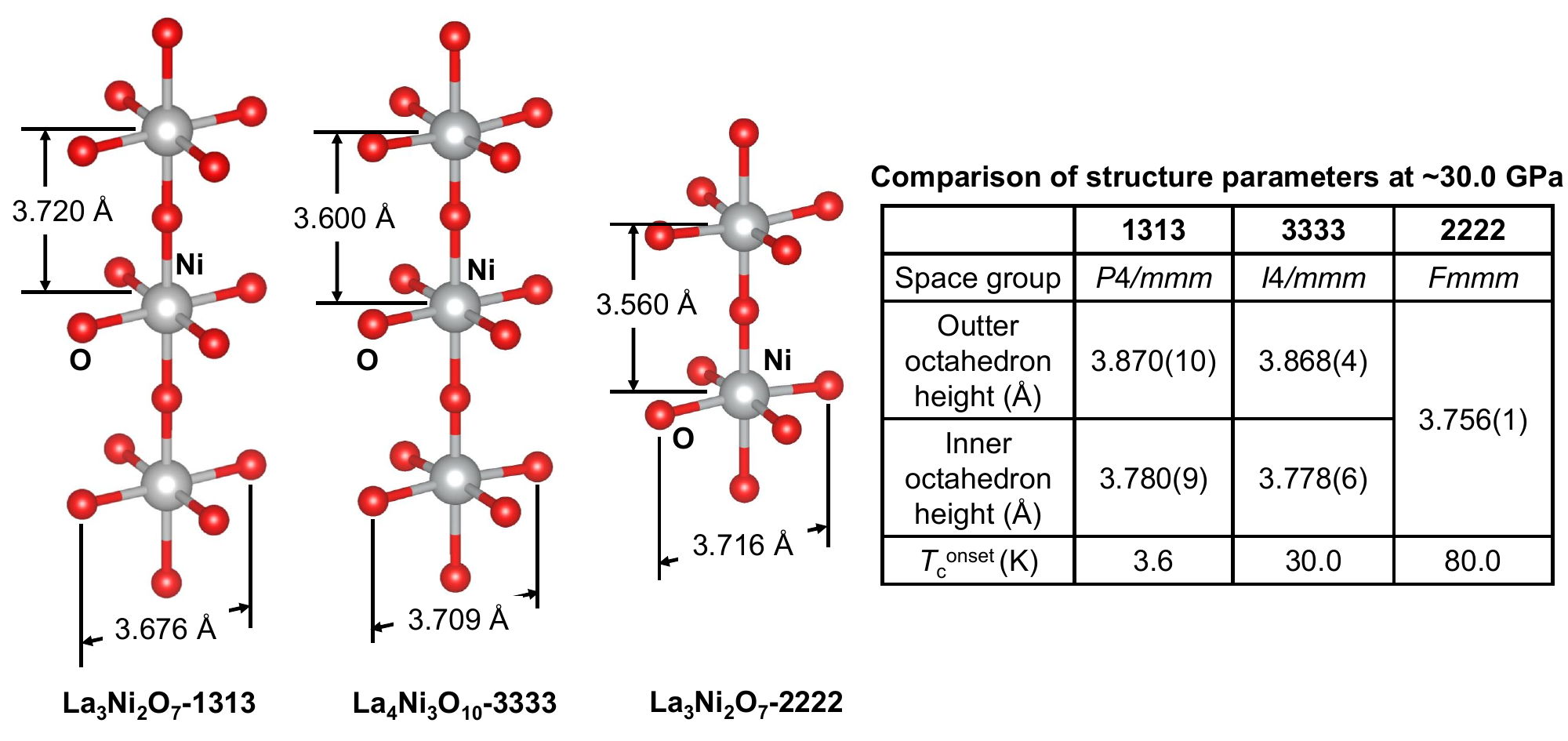}
    \caption*{\textbf{Extended Data Fig.3$|$Systematic analysis of the distances between adjacent Ni-cations of the 1313, 3333 and 2222 phase.}
    The left panel shows schemetics of the core trilayer/bilayer NiO$_6$ ochtahedron structures. Evidently, the 1313 phase exhibits the longest interlayer distance between two adjacent Ni cations. The 3333 and the 2222 phase share similar values\cite{li2025i,li2024str}. Table on the right panel summarizes detailed octahedron parameters. In-plane lattice constant $a$ of the 1313 phase is the smallest. Note that the in-plane lattic constant $a_p$ of the 2222 phase is calculated through $a_p$ = $(a + b)/2\sqrt{2}$.}
\vspace*{6in}
\end{figure*}

\end{document}